\documentclass[aps,prb,reprint,twocolumn,groupedaddress,showpacs]{revtex4}
\usepackage{amsmath}
\usepackage{amsfonts}
\usepackage{mathrsfs}
\usepackage{bm}
\usepackage{epsfig}

\begin{document}

\title{Conservation of Angular Momentum in a Flux Qubit}
\author{E. M. Chudnovsky, D. A. Garanin, and M. F. O'Keeffe}
\affiliation{Physics Department, Lehman College, City University
of New York,
    250 Bedford Park Boulevard West, Bronx, New York, 10468-1589, USA}

\date{\today}

\begin{abstract}
Oscillations of superconducting current between clockwise and
counterclockwise directions in a flux qubit do not conserve the
angular momentum of the qubit. To compensate for this effect the
solid containing the qubit must oscillate in unison with the
current. This requires entanglement of quantum states of the qubit
with quantum states of a macroscopic body. The question then
arises whether slow decoherence of quantum oscillations of the
current is consistent with fast decoherence of quantum states of a
macroscopic solid. This problem is analyzed within an exactly
solvable quantum model of a qubit embedded in an absolutely rigid
solid and for the elastic model that conserves the total angular
momentum. We show that while the quantum state of a flux qubit is,
in general, a mixture of a large number of rotational states, slow
decoherence is permitted if the system is macroscopically large.
Practical implications of entanglement of qubit states with
mechanical rotations are discussed.
\end{abstract}


\maketitle

\section{Introduction \label{Sec-intro}}

Flux qubits are formed by quantum superposition of current states
in a superconducting loop interrupted by one or more Josephson
junctions \cite{Leggett-Science02,Clarke-Nature08}. Quantum
mechanics of such a qubit is described by a double-well potential,
similar to the textbook example \cite{QM} of the ammonia molecule,
NH$_3$. In the latter example the tunneling between states
corresponding to the N-atom located to the left (L) or to the
right (R) of the H$_3$ triangle creates quantum superposition of
the $|L\rangle$ and $|R\rangle$ states, with the lowest energy
doublet given by $|L\rangle \pm |R\rangle$. If one prepares the
molecule in, e.g., the $|L\rangle$ state, the quantum mechanical
average of the position of the N-atom oscillates harmonically
between left and right at the frequency $\omega = \Delta/\hbar$,
where $\Delta$ is the energy splitting of the doublet. In the
rigorous formulation of this problem the N-atom and the H$_3$
triangle co-tunnel in such a manner that the position of the
center of mass of the four atoms is preserved, thus conserving the
linear momentum.

In the simplest formulation of the flux qubit problem the role of
left and right is played by clockwise and counterclockwise
directions of the current. Typical values of the angular momentum
associated with the current range from a few hundred $\hbar$ for a
submicron SQUID loop \cite{Wal-Science00}, to $10^5\hbar$ for a
micron-size loop \cite{Chiorescu-Science03}, to $10^{10}\hbar$ for
larger SQUIDs \cite{Friedman-Nature00}. To conserve the angular
momentum the tunneling of the current between clockwise and
counterclockwise directions must be accompanied by quantum
transitions between mechanical clockwise and counterclockwise
rotations of the body containing the flux qubit. This creates a
controversy \cite{Nikulov-10}. Indeed, the co-tunneling of the
superconducting current and mechanical rotation needed to conserve
the angular momentum requires entanglement of quantum states of
the flux qubit with quantum states of a macroscopic body. In any
reasonable experiment the phase of the wave function of the
equipment containing the flux qubit must be destroyed
instantaneously. Then how can the flux qubit preserve coherence on
a measurable time scale? This paper is devoted to the detailed
analysis of the entanglement of current states with mechanical
rotations and its implications for superconducting qubits.

Within an exactly solvable model of a flux qubit embedded in an
absolutely rigid rotator we obtain entangled eigenstates of the
system and their dependence on the total angular momentum $J$.
When the system is prepared in the state with a certain direction
of the superconducting current, this state is, in general, a
quantum mixture of many rotational states of the body. However,
only tunnel splitttings $\Delta_J$ of the states belonging to the
same $J$ contribute to the oscillations of the superconducting
current. We show that decoherence resulting from the broad
statistical distribution over $J$ is small as long as the body
containing the qubit is macroscopically large. Thus, contrary to
what one might think, the macroscopicity of the body that is
entangled with the qubit, is in fact required for low decoherence.
We then study decoherence of a flux qubit due to torques generated
by the oscillating current in the elastic solid and show how
decoherence rates obtained within the two models match. Among
other problems we discuss renormalization of the tunnel splitting
by the elastic environment and superradiant relaxation in a system
of closely packed qubits.

The paper is structured as follows. Exactly solvable quantum model
of a flux qubit interacting with rotations of a rigid body is
studied in \ref{sec-rigid}. Quantum states of the qubit entangled
with rotations of the body are obtained in Section
\ref{sub-rotations-rigid}. Section \ref{sub-decoherence-rigid} is
devoted to decoherence due to rotational excitations of the body.
Elastic environment is considered in Section \ref{sec-elastic}.
The model that conserves the total angular momentum is formulated
in Section \ref{sub-rotations-elastic}. Section
\ref{sub-decoherence-elastic} discusses decoherence of the flux
qubit by internal torques. Renormalization of the tunnel splitting
by the elastic environment is computed in Section
\ref{sub-renormalization}. Section \ref{sec-conclusion} contains
numerical estimates, discussion of various effects originating
from conservation of angular momentum, alternative interpretations
of the results, and final conclusions.

\section{Rigid Body \label{sec-rigid}}

\subsection{Rotational states of a flux qubit \label{sub-rotations-rigid}}

First, we consider the tunnel-split states of a flux qubit and
ignore conservation of the angular momentum. Let the lowest-energy
doublet of a flux qubit be
\begin{equation}\label{pm}
\Psi_{\pm} = \frac{1}{\sqrt{2}}\left(|\uparrow\rangle \pm
|\downarrow\rangle\right)\,,
\end{equation}
where $|\uparrow\rangle$ and $|\downarrow\rangle$ are the
eigenstates of the operator of the angular momentum of the
electronic current inside superconducting loop $\hat{l}_z$,
\begin{eqnarray}
\hat{l}_z|\uparrow\rangle & = & l|\uparrow\rangle \nonumber \\
\hat{l}_z|\downarrow\rangle & = & -l|\downarrow\rangle
\end{eqnarray}
Eigenfunctions $\Psi_{\pm}$ satisfy
\begin{equation}\label{Epm}
\hat{H}\Psi_{\pm} = E_{\pm}\Psi_{\pm}
\end{equation}
with $\hat{H}$ being the Hamiltonian of the qubit and
\begin{equation}\label{delta}
E_- - E_+ \equiv \Delta
\end{equation}
being the tunnel splitting. It is convenient to describe such a
two-state system by a pseudospin 1/2. Components of the
corresponding Pauli operator ${\bm \sigma}$ are
\begin{eqnarray}\label{Pauli}
\sigma_x & = & |\downarrow\rangle\langle \uparrow| +
|\uparrow\rangle\langle \downarrow|
 \nonumber \\
 \sigma_y & = &
i|\downarrow\rangle\langle \uparrow| - i|\uparrow\rangle\langle
\downarrow| \nonumber \\ \sigma_z & = & |\uparrow\rangle\langle
\uparrow| - |\downarrow\rangle\langle \downarrow|\,.
\end{eqnarray}
The projection of $\hat{H}$ onto $|\uparrow\rangle$ and
$|\downarrow\rangle$ states is
\begin{equation}\label{projection}
\hat{H}_{\sigma} = \sum_{m,n = \uparrow,\downarrow}\langle
m|\hat{H}|n\rangle|m\rangle\langle n|\,.
\end{equation}
According to Eq.\ (\ref{pm}),
\begin{eqnarray}
&& |\uparrow\rangle = \frac{1}{\sqrt{2}}\left(\Psi_{+} + \Psi_{-}\right) \nonumber \\
&& |\downarrow\rangle = \frac{1}{\sqrt{2}}\left(\Psi_{+} -
\Psi_{-}\right)\,.
\end{eqnarray}
It is now easy to see from Eq.\ (\ref{Epm}) that
\begin{eqnarray}\label{ME}
& & \langle \uparrow|\hat{H}|\uparrow\rangle = \langle
\downarrow|\hat{H}|\downarrow\rangle = 0 \nonumber \\
& & \langle \downarrow| \hat{H}|\uparrow\rangle = \langle
\uparrow| \hat{H}|\downarrow\rangle = - {\Delta}/{2}\,.
\end{eqnarray}
With the help of these relations one obtains from equations
(\ref{Pauli}) and (\ref{projection})
\begin{equation}\label{projection1}
\hat{H}_{\sigma} = -({\Delta}/{2})\sigma_x\,.
\end{equation}

The general form of the wave function of our two-state system is
\begin{equation}
\Psi(t) =  C_+ \Psi_+ e^{i\Delta t/(2\hbar)} +
C_-\Psi_-e^{-i\Delta t/(2\hbar)}
\end{equation}
with $|C_-|^2 + |C_+|^2 = 1$. If one imposes the initial condition
$\Psi(0) = |\uparrow\rangle$, then
\begin{equation}
\Psi(t) = \cos\left(\frac{\Delta t}{2\hbar}\right)|\uparrow
\rangle + \sin\left(\frac{\Delta t}{2\hbar}\right)|\downarrow
\rangle
\end{equation}
and $\langle \hat{l}_z \rangle = l \langle \sigma_z \rangle$, with
\begin{equation}\label{oscillations}
\langle \sigma_z \rangle = \langle \Psi(t)|\sigma_z|\Psi(t)\rangle
= \cos\left(\frac{\Delta t}{\hbar}\right)\,.
\end{equation}
This equation describes harmonic oscillations of the
superconducting current at the frequency $\Delta/\hbar$ between
clockwise and counterclockwise directions. Another way to obtain
this result is to use the equivalence \cite{lectures} of the
Schr\"{o}dinger equation for spin one-half to the precession
equation for the expectations value of ${\bm \sigma}$,
\begin{equation}
\hbar \frac{d}{dt}\left\langle\frac{\bm \sigma}{2}\right\rangle =
-\left\langle {\bm \sigma} \times
\frac{\delta\hat{H}_{\sigma}}{\delta {\bm \sigma}}\right\rangle =
\frac{\Delta}{2} \langle {\bm \sigma} \rangle \times {\bf e}_x \,,
\end{equation}
which gives
\begin{eqnarray}
&& \frac{d}{dt}\langle \sigma_x \rangle = 0 \nonumber \\
&& \frac{d}{dt}\langle \sigma_y \rangle =  \frac{\Delta}{\hbar}
\langle \sigma_z \rangle \nonumber \\
&& \frac{d}{dt}\langle \sigma_z \rangle =  -\frac{\Delta}{\hbar}
\langle \sigma_y \rangle\,.
\end{eqnarray}
The last two equations give Eq.\ (\ref{oscillations}).

We shall account now for mechanical rotations of the body
containing the flux qubit. In this Section we shall deal with an
absolutely rigid body that can only rotate as a whole. As we shall
see, this problem contains all of the components needed to
understand the effects of entanglement required by the
conservation of the angular momentum. Rotation by the angle $\phi$
about the quantization axis $Z$ transforms the Hamiltonian of the
qubit into
\begin{equation}\label{spinrotation}
\hat{H}'  = e^{-i\hat{l}_z\phi}\hat{H} e^{i\hat{l}_z\phi}\,.
\end{equation}
Noticing that the operator of the angular momentum of the
superconducting current, $\hat{l}_z$ (that is chosen in units of
$\hbar$), commutes with $\phi$ it is easy to project this
Hamiltonian onto $|\uparrow\rangle$ and $|\downarrow\rangle$.
Simple calculation yields the following generalization of Eq.\
(\ref{projection1}):
\begin{eqnarray}\label{projection'}
\hat{H}_{\sigma}' & = & \sum_{m,n = \uparrow,\downarrow}\langle
m|\hat{H}'|n\rangle|m\rangle\langle n|   \nonumber \\
&=& -\frac{\Delta}{2}\left[e^{-2il\phi}\sigma_+ +
e^{2il\phi}\sigma_-\right] \nonumber \\ & = &
-\frac{\Delta}{2}\left[\cos(2l\phi)\sigma_x +
\sin(2l\phi)\sigma_y\right]   \,
\end{eqnarray}
where $\sigma_{\pm} = \frac{1}{2}(\sigma_x \pm i\sigma_y)$.

To develop a rigorous formulation of the problem let us first
assume that the body with the qubit is an isolated system in a
pure quantum state described by a single wave function. The full
Hamiltonian of the system is
\begin{equation}\label{H-phi}
\hat{H} = \frac{(\hbar \hat{L}_z)^2}{2I} -\frac{\Delta}{2}\left[
\sigma_x \cos(2l\phi)+ \sigma_y \sin(2l\phi)\right]\,,
\end{equation}
where $\hat{L}_z = -i(d/{d \phi})$ and $I \equiv I_z$ is the
moment of inertia of the body for rotation about the quantization
axis. It is easy to check that this Hamiltonian commutes with the
operator of the total angular momentum,
\begin{equation}
\hat{J}_z = \hat{L}_z + \hat{l}_z = -i\frac{d}{d\phi} +
l\sigma_z\,.
\end{equation}
Consequently, the eigenstates of (\ref{H-phi}) must be entangled
states of $\hat{l}_z$ and $\hat{L}_z$ that are eigenstates of the
total angular momentum $\hat{J}_z$:
\begin{equation}\label{Psi-J}
|\Psi_{J\pm}\rangle =
\frac{C_{J\pm}}{\sqrt{2}}|\uparrow\rangle_l\otimes|J-l\rangle_{L}
\pm \frac{C_{J\mp}}{\sqrt{2}}|\downarrow\rangle_l\otimes|J
+l\rangle_L \,,
\end{equation}
with $J \equiv J_z$. Simple calculation gives
\begin{equation}
C_{J\pm} = \sqrt{ 1 \pm \frac{1}{\sqrt{1 +
\frac{\Delta^2I^2}{4(\hbar l)^2(\hbar J)^2}}}} \label{eq:C}
\end{equation}
and
\begin{equation}\label{levels}
E_{J\pm} = \frac{(\hbar l)^2}{2I} + \frac{(\hbar J)^2}{2I} \pm
\sqrt{\frac{\Delta^2}{4} + \frac{(\hbar l)^2(\hbar J)^2}{I^2}}
\end{equation}
for the energy levels. Here $\pm$ corresponds to $\mp$ in Eq.\
(\ref{Psi-J}) and $J = 0, \pm 1, \pm 2, ...$.

Alternatively, the same results can be obtained in the coordinate
frame attached to the current loop. In this case one starts with
the Hamiltonian
\begin{equation}
\hat{H}_r = \frac{(\hbar \hat{L}_z)^2}{2I}
-\frac{\Delta}{2}\sigma_x = \frac{(\hbar \hat{J}_z - \hbar
\hat{l_z} )^2}{2I} -\frac{\Delta}{2}\sigma_x \,.
\end{equation}
Its eigenfunctions are
\begin{equation}
|\Psi_{J\pm}\rangle_r =
\frac{1}{\sqrt{2}}\left(C_{J\pm}|\uparrow\rangle_l \pm
C_{J\mp}|\downarrow\rangle_l\right) \otimes|J\rangle\,,
\end{equation}
while eigenvalues are given by Eq.\ (\ref{levels}). The two
coordinate frames are related by  unitary transformation.

\subsection{Decoherence from rotations \label{sub-decoherence-rigid}}

Any real macroscopic system should have some distribution over
$J$. According to Eq.\ (\ref{levels}), at large $I$ the energies
of the states corresponding to different $J$ can be very close.
Consequently, a macroscopically large number of different
$J$-states should contribute to the expectation value of any
physical quantity. Since the phases of such states can differ
significantly, the question then arises how the coherence of the
flux qubit is influenced by this effect. Rigorous answer to this
question is given below.

To study decoherence, one should prepare the system in a state
with a certain direction of $l_z$, e.g. $l_z = +l$, and study how
$\langle \hat{l}_z \rangle$ would depend on time. Naturally, the
initial state should be obtained by subjecting the system to a
strong bias field in the direction of the magnetic moment of the
current loop. Adding the term $-\frac{1}{2}W\sigma_z$ to the
Hamiltonian, it is easy to work out the energy levels of the
biased states:
\begin{equation}
E_{J\pm} = \frac{(\hbar l)^2}{2I} + \frac{(\hbar J)^2}{2I} \pm
\sqrt{\frac{\Delta^2}{4} + \left[\frac{W}{2} + \frac{(\hbar
l)(\hbar J)}{I}\right]^2}\,.
\end{equation}
For a large positive bias the states corresponding to the plus
sign in the above equation have too high energies and can be
ignored. In this limit the relevant energies, up to a constant,
are
\begin{equation}\label{E-J}
E_{J-} \equiv E_J = \frac{\hbar^2(J-l)^2}{2I} = \frac{(\hbar
L_z)^2}{2I}\,,
\end{equation}
in accordance with the expectation that they must be the energies
of the rotational states of the body. To make sure that the system
is magnetized in the direction of the field, that is $l_z = +l$,
it must be put in contact with a thermal bath at temperature $T$.
This provides thermal distribution over $E_J$ with probabilities
given by
\begin{equation}\label{A}
P_J = \frac{1}{Z}\exp\left(-\frac{E_J}{k_BT}\right)\,, \qquad Z =
\sum_J\exp\left(-\frac{E_J}{k_BT}\right)\,.
\end{equation}
If at $t = 0$ the field is removed and the system is isolated from
the bath, it will be a mixture of $J$-states,
\begin{equation}\label{initial}
|\Psi_{Jl}\rangle_0 = |l\rangle \otimes |J-l\rangle\,,
\end{equation}
with the probability of each $J$ determined by Eq.\ (\ref{A}).
Time evolution of each $J$-state is provided by
\begin{equation}\label{Psi-J-l}
|\Psi_{Jl}\rangle =\frac{C_{J+}}{\sqrt{2}}|\Psi_{J+}\rangle
e^{-iE_{J+}t/\hbar} + \frac{C_{J-}}{\sqrt{2}}|\Psi_{J-}\rangle
e^{-iE_{J-}t/\hbar}\,.
\end{equation}
Consequently, the time dependence of the expectation value of
$\hat{l}_z = l\sigma_z$ is determined by
\begin{equation}
\langle \sigma_z \rangle  = \sum_J P_J \langle
\Psi_{Jl}|\sigma_z|\Psi_{Jl}\rangle\,.
\end{equation}
Using the relations
\begin{eqnarray}
\langle \Psi_{J+}|\sigma_z|\Psi_{J+}\rangle & = &
\frac{1}{2}\left(C^2_{J+} - C^2_{J-}\right) \nonumber \\
\langle \Psi_{J-}|\sigma_z|\Psi_{J-}\rangle & = &
\frac{1}{2}\left(C^2_{J-} - C^2_{J+}\right) \nonumber \\
\langle \Psi_{J-}|\sigma_z|\Psi_{J+}\rangle & = & \langle
\Psi_{J+}|\sigma_z|\Psi_{J-}\rangle = C_{J+}C_{J-}
\end{eqnarray}
one obtains
\begin{equation}\label{sum-J}
\langle \sigma_z \rangle = \sum_{J}P_J\left[\frac{\beta^2_J}{1 +
\beta^2_J} + \frac{1}{1 +
\beta^2_J}\cos\left(\frac{\Delta_J}{\hbar}t\right)\right]\,,
\end{equation}
where
\begin{equation}\label{Delta-J}
\Delta_J = E_{J+} - E_{J-} = \Delta\sqrt{1 + \beta^2_J}\,, \qquad
\beta_J = \frac{2(\hbar l)(\hbar J)}{I\Delta}\,.
\end{equation}
Notice that only the energy splitting between states belonging to
the same $J$, separated by $\Delta_J$, contribute to $\langle
\sigma_z \rangle$. For a given $J \neq 0$ oscillations of the
superconducting current occur between $\langle l_z \rangle = l$
and $\langle l_z \rangle = l(\beta^2_J - 1)/(\beta^2_J + 1)$ as
compared to the oscillations between $\pm l$ for $J = 0$ ($\beta_J
= 0$).

Formally, at $T = 0$, only the non-rotating state with $J = l$
contributes to the sum in Eq.\ (\ref{sum-J}), providing
\begin{equation}\label{T=0}
\langle \sigma_z \rangle = \frac{\beta^2_l}{1 + \beta^2_l} +
\frac{1}{1 + \beta^2_l}\cos\left(\frac{\Delta_l}{\hbar}t\right)\,,
\end{equation}
where $\beta_l$ equals $\beta_J$ at $J = l$. For a macroscopic
body with a large moment of inertia $\beta_l \ll 1$, so that the
difference between Eq.\ (\ref{oscillations}) and Eq.\ (\ref{T=0})
is very small. The absence of decoherence at $T = 0$ is related to
the fact that the system is in a pure $J$-state.

At $T \neq 0$ rotations of a macroscopic body must be distributed
over a macroscopically large number of $J \gg l$. Consequently,
one can replace $J-l$ in Eq.\ (\ref{E-J}) with $J$ and replace
summation in Eqs.\ (\ref{sum-J}), (\ref{A}) by integration over
$J$. This gives $Z = \sqrt{2\pi I k_B T}/\hbar$. Expectation value
of $\sigma_z$ depends on time through $(\Delta/\hbar)t$,
\begin{eqnarray}\label{int-beta}
&& \langle \sigma_z \rangle =
\frac{1}{\sqrt{\pi}\beta_T}\int_{-\infty}^{+\infty} d\beta_J
\exp\left(-\frac{\beta_J^2}{\beta_T^2}\right) \times \nonumber
\\
&&\left[\frac{\beta^2_J}{1 + \beta^2_J} + \frac{1}{1 +
\beta^2_J}\cos\left(\sqrt{1 + \beta_J^2}\frac{\Delta
}{\hbar}t\right)\right]\,,
\end{eqnarray}
and is determined by a single parameter,
\begin{equation}\label{beta-T}
\beta_T = 2\sqrt{\beta_l\frac{k_B T}{\Delta}} = 2^{3/2}\frac{\hbar
l}{\Delta}\sqrt{\frac{k_B T}{I}} \,.
\end{equation}
Note that $\beta_l = 2(\hbar l)^2/(I\Delta)$ contains a
macroscopically large number $I$ in the denominator. This provides
\begin{equation}
\beta_l \ll \beta_T \ll 1
\end{equation}
for any reasonable values of $l$, $\Delta$, and $T$. Since the
main contribution to the integral in Eq.\ (\ref{int-beta}) comes
from $\beta_J \sim \beta_T \gg \beta_l$, the overwhelming majority
of $J$ contributing to the integral satisfy $J \gg l$ in
accordance with our assumption.

From Eq.\ (\ref{int-beta}) the asymptotic value of $\langle
\sigma_z \rangle$ is
\begin{equation}
\sigma_{\infty} \equiv \lim_{t \rightarrow \infty} \langle
\sigma_z \rangle =\frac{1}{2}\beta_T^2 = 2\beta_l\frac{k_B
T}{\Delta}\,.
\end{equation}
For a macroscopic body it is small due to the smallness of
$\beta_l$. In this limit the time dependence of the oscillating
term in Eq.\ (\ref{int-beta}) can be computed exactly:
\begin{equation}\label{Re}
\langle \sigma_z \rangle_t = {\rm
Re}\left[\frac{e^{i(\Delta/\hbar)t}}
{\sqrt{1-i\sigma_{\infty}(\Delta/\hbar)t}}\right]\,.
\end{equation}
One can see that the amplitude of quantum oscillations is
decreasing as $1/\sqrt{\sigma_{\infty}(\Delta/\hbar)t}$. Thus, the
effective decoherence rate due to the entanglement of the flux
qubit with rotations of the rigid body is
\begin{equation}\label{rate-rigid}
\Gamma_r = \sigma_{\infty}\frac{\Delta}{\hbar} = 2\beta_l\frac{k_B
T}{\hbar} = \frac{4\hbar l^2}{I}\left(\frac{k_B
T}{\Delta}\right)\,.
\end{equation}
Notice that slow, $1/\sqrt{t}$, decay of coherent oscillations
given by Eq.\ (\ref{Re}) is a consequence of the absolute rigidity
of the body.

Proportionality of $\Gamma_r$ to $1/I$ illustrates our point that,
contrary to the naive picture that one might have
\cite{Nikulov-10}, the entanglement of a flux qubit with rotations
of a macroscopic body, dictated by the conservation of angular
momentum, does not necessarily result in a strong decoherence.
This comes as a consequence of the selection rule: According to
Eq.\ (\ref{sum-J}) only tunnel splittings, $\Delta_J = E_{J+} -
E_{J-}$, of the states (\ref{levels}) belonging to the same $J$
contribute to $\langle \sigma_z \rangle$. For a macroscopic body,
all $\Delta_J$ are very close, thus providing low decoherence.

\section{Elastic Body \label{sec-elastic}}

\subsection{Flux qubit in
the elastic environment \label{sub-rotations-elastic}}

Realistically, the body containing a flux qubit is not absolutely
rigid. During half-period of oscillations of the superconducting
current the elastic stress generated by the changing angular
momentum of the current may only extend as far as half-wavelength,
$\lambda/2 = \pi\hbar v_t/\Delta$, of the transverse sound of
frequency $\Delta/\hbar$ and speed $v_t$. We shall assume that
this distance is greater than the size of the current loop. For,
e.g., a micron-size loop this condition would be typically
fulfilled for $\Delta/\hbar < 10$GHz. It allows one to treat the
flux qubit as a point source of the elastic stress, without
considering interactions of segments of the current loop with the
elastic environment.

Now the rotation angle $\phi$ that appears in the previous section
is determined to the elastic twist, \cite{LL-elasticity}
\begin{equation}\label{phi}
\phi = \frac{1}{2}[{\bm \nabla} \times {\bf u}]_z\,,
\end{equation}
where ${\bf u}$ is the phonon displacement field at the location
of the flux qubit ${\bf r} = 0$. Conventional quantization of
phonons gives
\begin{equation}\label{phi-quantized}
\phi = \frac{1}{2}\sqrt{\frac{\hbar }{2\rho V}}\sum_{\mathbf{k}%
\lambda }\frac{\left[ i\mathbf{k}\times \mathbf{e}_{\mathbf{k}\lambda }%
\right]_z}{\sqrt{\omega _{\mathbf{k}\lambda }}}\left(
a_{\mathbf{k}\lambda }+a_{-\mathbf{k}\lambda }^{\dagger
}\right)\,,
\end{equation}
where $a_{\mathbf{k}\lambda }^{\dagger}, a_{\mathbf{k}\lambda }$
are operators of creation and annihilation of phonons of
wave-vector ${\bf k}$ and polarization $\lambda$,
$\mathbf{e}_{\mathbf{k}\lambda }$ are unit vectors of
polarization, $\omega _{\mathbf{k}\lambda } = v_tk$ is the phonon
frequency, $\rho$ is the mass density of the solid and $V$ is its
volume. Since we limit our consideration to elastic twists, only
the two transverse polarizations of sound contribute to Eq.\
(\ref{phi-quantized}).

Elastic Hamiltonian that replaces Hamiltonian (\ref{H-phi}) of the
rigid-body approximation is
\begin{eqnarray}\label{SQUID-phonons}
\hat{H} & = & \sum_{\mathbf{k} \lambda } \hbar \omega
_{\mathbf{k}\lambda }\left(a_{\mathbf{k}\lambda
}^{\dagger}a_{\mathbf{k}\lambda } +\frac{1}{2}\right) -\nonumber
\\
 & - & \frac{\Delta}{2}\left\{\sigma_+
\exp\left[l\sum_{\mathbf{k}\lambda }\xi_{{\mathbf{k}\lambda
}}\left( a_{\mathbf{k}\lambda }-a_{\mathbf{k}\lambda }^{\dagger
}\right)\right]  \right. \nonumber \\
& + & \left. \sigma_- \exp\left[ -l\sum_{\mathbf{k} \lambda
}\xi_{{\mathbf{k}\lambda }}\left( a_{\mathbf{k}\lambda
}-a_{\mathbf{k}\lambda }^{\dagger }\right)\right]\right\} \,,
\end{eqnarray}
where
\begin{equation}\label{xi}
\xi_{{\mathbf{k}\lambda }} \equiv \sqrt{\frac{\hbar }{2\rho
V}}\frac{\left[ \mathbf{k}\times \mathbf{e}_{\mathbf{k}\lambda }
\right]_z}{\sqrt{\omega _{\mathbf{k}\lambda }}}\,.
\end{equation}
Validity of this approximation relies on the fact that angular
velocity of the local rotation, $\Omega = d\phi/dt$, is always
small compared to the frequency of sound $\omega$. Indeed,
noticing that according to Eq.\ (\ref{phi}) $\Omega \sim \omega
ku$ we see that $\Omega \ll \omega$ coincides with the condition
of validity of the elastic theory: $ku \ll 1$.

Unitary transformation $\hat{H}_r = \hat{U}^{-1}\hat{H}\hat{U}$
with
\begin{equation}
\hat{U} = \exp\left[\frac{1}{2}\,l\sigma_z\sum_{\mathbf{k} \lambda
}\xi_{{\mathbf{k}\lambda }}\left( a_{\mathbf{k}\lambda
}-a_{\mathbf{k}\lambda }^{\dagger }\right)\right]
\end{equation}
transforms Hamiltonian (\ref{SQUID-phonons}) into
\begin{eqnarray}\label{H-rot}
&& \hat{H}_r  =  \hat{U}^{-1}\left[\sum_{\mathbf{k} \lambda }
\hbar \omega _{\mathbf{k}\lambda }\left(a_{\mathbf{k}\lambda
}^{\dagger}a_{\mathbf{k}\lambda }
+\frac{1}{2}\right)\right]\hat{U} - \frac{\Delta}{2}\sigma_x
\nonumber \\
&& =\sum_{\mathbf{k} \lambda }\hbar \omega _{\mathbf{k}\lambda }
\left[a_{\mathbf{k}\lambda }^{\dagger}a_{\mathbf{k}\lambda }  -
\frac{l\sigma_z}{2}\xi_{{\mathbf{k}\lambda }}\left(
a_{\mathbf{k}\lambda }+a_{\mathbf{k}\lambda }^{\dagger
}\right)\right] -
\frac{\Delta}{2}\sigma_x \,, \nonumber \\
\end{eqnarray}
where an insignificant constant has been omitted. In the
transition from the first to the second line of Eq.\ (\ref{H-rot})
we have used properties of the displacement operator,
\begin{equation}
\hat{D}^{-1}(\alpha)a\hat{D}(\alpha) = a + \alpha\,, \quad
\hat{D}^{-1}(\alpha)a^{\dagger}\hat{D}(\alpha) = a^{\dagger} +
\alpha^*\,,
\end{equation}
with
\begin{equation}
\hat{D}(\alpha_{\mathbf{k}\lambda }) =
e^{-\alpha_{\mathbf{k}\lambda }^* a_{\mathbf{k}\lambda } +
\alpha_{\mathbf{k}\lambda } a_{\mathbf{k}\lambda
}^{\dagger}}\,,\quad \alpha_{\mathbf{k}\lambda } =
-\frac{1}{2}\,l\sigma_z\xi_{{\mathbf{k}\lambda }}\,.
\end{equation}

Eq.\ (\ref{H-rot}) shows that from mathematical point of view the
problem formulated in this Section is a variance of spin-boson
problem \cite{Leggett-spinboson}. While some important theorems
have been proved for this problem in recent years (see, e.g.,
Ref.\ \onlinecite{Gardas-JPhys11} and references therein), its
exact eigenstates are unknown. This prevents us from developing
rigorous mathematical approach to decoherence along the lines of
the previous Section. From a physical point of view, the
attractiveness of our variance of the spin-boson model is in the
absence of free parameters. The boson field in our case is the
phonon displacement field. Its coupling to the flux qubit
(described by spin 1/2) is completely determined by the
conservation of total angular momentum. In what follows, we will
use an approximation based upon observation that local twists of
the elastic solid due to oscillations of the superconducting
current in a flux qubit must be very small. Within this
approximation we will describe transverse phonons by a classical
displacement field ${\bf u}({\bf r},t)$, satisfying ${\bm
\nabla}\cdot {\bf u} = 0$.

Expanding Hamiltonian (\ref{SQUID-phonons}) to the lowest power on
the elastic twist and replacing operators by their classical
expectation values, one obtans
\begin{equation}\label{expansion}
{H} = {H}_E -\frac{\Delta}{2} \sigma_x -
\frac{\Delta}{2}l\sigma_y\int d^3r \delta({\bf r})
\left(\frac{\partial u_x}{\partial y} - \frac{\partial
u_y}{\partial x}\right)\,,
\end{equation}
where ${H}_E$ is the Hamiltonian of free rotations,
\begin{equation}
{H}_E =  \frac{1}{4}\int d^3r \rho v_t^2\left(\frac{\partial
u_{\alpha}}{\partial r_{\beta}} + \frac{\partial
u_{\beta}}{\partial r_{\alpha}}\right)^2\,.
\end{equation}
The dynamical equation for the displacement field is
\begin{equation}\label{elastic}
\rho \frac{\partial^2 u_{\alpha}}{\partial t^2} = \frac{\partial
\sigma_{\alpha \beta}}{\partial r_{\beta}} \,,
\end{equation}
where $\sigma_{\alpha \beta} = {\delta {H}}/\delta e_{\alpha
\beta}$ is the stress tensor and $e_{\alpha \beta} =
\partial u_{\alpha}/\partial r_{\beta}$ is the strain tensor.
This gives
\begin{eqnarray}
&& \rho \left(\frac{\partial^2 u_{x}}{\partial t^2} -v_t^2{
\nabla}^2u_x \right) = - \frac{\Delta}{2}l\sigma_y \frac{\partial
}{\partial
y}\delta({\bf r}) \label{ux} \nonumber \\
\\
&& \rho \left(\frac{\partial^2 u_{y}}{\partial t^2}
-v_t^2\nabla^2u_y\right) = \frac{\Delta}{2}l\sigma_y
\frac{\partial
}{\partial x}\delta({\bf r})\label{uy}\,. \nonumber \\
\end{eqnarray}
The above equations should be solved together with the
Landau-Lifshitz equation for ${\bm \sigma}$:
\begin{equation}\label{LL-sigma}
\frac{\hbar}{2} \frac{d{\bm \sigma}}{d t} = -{\bm \sigma} \times
\frac{\delta {H}}{\delta {\bm \sigma}}\,,
\end{equation}
which gives
\begin{eqnarray}
&&\hbar\frac{d \sigma_x}{d t} = -\sigma_z\Delta l \int d^3 r
\delta({\bf r})\left[\frac{\partial u_x}{\partial y} -
\frac{\partial u_y}{\partial x}\right] \label{LLx}\\
&&\hbar\frac{d \sigma_y}{d t} = \sigma_z \Delta \label{LLy}\\
&&\hbar\frac{d \sigma_z}{d t} = -\sigma_y \Delta + \sigma_x\Delta
l \int d^3 r \delta({\bf r})\left[\frac{\partial u_x}{\partial y}
- \frac{\partial u_y}{\partial x}\right]\label{LLz}
\end{eqnarray}
It is easy to see that Eqs.\ (\ref{LLx}), (\ref{LLy}) and
(\ref{LLz}) preserve the length of ${\bm \sigma}$: $\sigma_x^2 +
\sigma_y^2 + \sigma_z^2 = 1$.

First, let us show that, in accordance with our general line of
reasoning, the above equations conserve the $Z$-component of the
total angular momentum,
\begin{equation}
J_z = \hbar l\sigma_z + L_z\,.
\end{equation}
Here $L_z$ is the $Z$-component of the mechanical angular
momentum. Its time derivative equals the $Z$-component of the
total mechanical torque, $K_z$, acting on the body. In the absence
of the external torque applied to the surface of the body, $K_z$
is given by \cite{LL-elasticity}
\begin{equation}\label{torque}
{K}_{z} = \int d^3r \, (\sigma_{yx} - \sigma_{xy})\,.
\end{equation}
Conventional elastic theory postulates no internal torques, in
which case the stress tensor would be symmetric and $K_z$ would be
zero. Situation changes when there are transitions between angular
momentum states of a microscopic object inside the body, such as,
e.g., a flux qubit. In this case the stress tensor is
non-symmetric, yielding
\begin{equation}
\frac{d {L}_{z}}{dt} = \int d^3r \, (\sigma'_{yx} -
\sigma'_{xy})\,,
\end{equation}
where $\sigma'_{\alpha \beta} = \delta {H}_{\rm int}/\delta
e_{\alpha \beta}$ is the part of the stress tensor related to the
interaction of the flux qubit with the elastic environment,
$H_{\rm int}$. The latter is given by the second term in Eq.\
(\ref{SQUID-phonons}). To prove conservation of the total angular
momentum one needs to write this term with the accuracy to
second-order terms on the elastic twists:
\begin{eqnarray}
H_{\rm int} & = & -\frac{l}{2}\Delta \sigma_y\int d^3r \delta({\bf
r}) \left(\frac{\partial u_x}{\partial y} - \frac{\partial
u_y}{\partial x}\right) \nonumber \\
& + & \frac{l^2}{4}\Delta \sigma_x\left[\int d^3r \delta({\bf r})
\left(\frac{\partial u_x}{\partial y} - \frac{\partial
u_y}{\partial x}\right)\right]^2\,.
\end{eqnarray}
This gives
\begin{eqnarray}
\sigma'_{xy} & = & -\frac{l}{2} \Delta \sigma_y \delta({\bf r})+
\frac{l^2}{2}\Delta \sigma_x \delta({\bf r})\int d^3r \delta({\bf
r}) \left[\frac{\partial u_x}{\partial y} - \frac{\partial
u_y}{\partial x}\right] \nonumber \\
\sigma'_{yx} & = & \frac{l}{2} \Delta \sigma_y \delta({\bf
r})-\frac{l^2}{2}\Delta \sigma_x \delta({\bf r})\int d^3r
\delta({\bf r}) \left[\frac{\partial u_x}{\partial y} -
\frac{\partial u_y}{\partial x}\right] \nonumber \\
\end{eqnarray}
so that
\begin{eqnarray}\label{torque-z}
\frac{d{J}_z}{dt} & = & \hbar l\frac{d{\sigma}_z}{dt} +
\frac{d{L}_{z}}{dt}  =  \hbar l\frac{d{\sigma}_z}{dt} +
l\Delta\sigma_y \nonumber \\
& - & l^2\Delta\sigma_x \int d^3r \delta({\bf r})
\left(\frac{\partial u_x}{\partial y} - \frac{\partial
u_y}{\partial x}\right)\,.
\end{eqnarray}
It is now easy to see that condition $d{J}_z/dt = 0$ coincides
with one of the equations of motion, Eq.\ (\ref{LLz}).

\subsection{Decoherence from internal torques
\label{sub-decoherence-elastic}}

At ${\bf u} = 0$ equations (\ref{LLy}) and (\ref{LLz}) would
describe coherent precession of ${\bm \sigma}$ about the $X$-axis,
with $\sigma_x = const$, $\sigma_{z} \propto \cos(t\Delta/\hbar)$,
and $\sigma_{y} \propto \sin(t\Delta/\hbar)$. Conservation of
angular momentum makes the flux qubit wiggle mechanically when the
current oscillates between clockwise and counterclockwise.
Consequently, it becomes a source of sound, as can be seen from
Eqs.\ (\ref{ux}) and (\ref{uy}). Let us linearize all equations of
motion around $\sigma_x = 1$, ${\bf u} = 0$, with small
$\sigma_{y,z}(t) \propto e^{-i\omega t}$ and
\begin{equation}
u_{x,y}({\bf r},t) \propto e^{-i\omega
t}\int\frac{d^3k}{(2\pi)^3}e^{i{\bf k}\cdot{\bf r} }u_{x,y}({\bf
k} )\,.
\end{equation}
Writing $\delta(\bf r)$ as $\int\frac{d^3k}{(2\pi)^3}e^{i{\bf
k}\cdot{\bf r}}$ one obtains from Eqs. (\ref{ux}) and (\ref{uy})
\begin{equation}\label{u-k}
u_x({\bf k}) =
-\frac{l\Delta}{2\rho}\,\frac{ik_y\sigma_y}{k^2v_t^2 -
\omega^2}\,, \quad u_y({\bf k}) =
\frac{l\Delta}{2\rho}\,\frac{ik_x\sigma_y}{k^2v_t^2 - \omega^2}\,,
\end{equation}
where $k^2 = k_x^2 + k_y^2 + k_z^2$. Substitution into Eqs.
(\ref{LLy}) and (\ref{LLz}) results in
\begin{equation}\label{dispersion}
\hbar^2\omega^2 = \Delta^2 \left(1-\frac{l^2\Delta}{2\rho}\int
\frac{d^3k}{(2\pi)^3}\,\frac{k_x^2 + k_y^2}{k^2v_t^2 -
\omega^2}\right)
\end{equation}
The integral in this equation should be computed in the complex
plane with account of a small imaginary part of $\omega$,
\begin{equation}
\int \frac{d^3k}{(2\pi)^3}\,\frac{k_x^2 + k_y^2}{k^2v_t^2 -
\omega^2}  = \frac{1}{3\pi^2}\int\frac{k^4 dk}{k^2v_t^2 -
\omega^2} = \frac{i\omega^3}{3\pi v_t^5}\,.
\end{equation}
This gives
\begin{equation}
\hbar^2\omega^2 = \Delta^2 \left(1-i\frac{l^2\omega^3\Delta}{6\pi
\rho v_t^5}\right)\,,
\end{equation}
that is,
\begin{equation}
\omega = \frac{\Delta}{\hbar} - i\Gamma_0\,,
\end{equation}
where
\begin{equation}\label{Gamma}
\Gamma_0 = \frac{l^2\Delta^5}{12\pi \hbar^4 \rho v_t^5}
\end{equation}
is the $T = 0$ rate of the decay of the coherent precession of
${\bm \sigma}$. This result is in full agreement with the
decoherence rate computed with the help of the Fermi golden rule
by considering spontaneous quantum transition from the excited
state $\left(|l\rangle - |-l\rangle\right)$ to the ground state
$\left(|l\rangle + |-l\rangle\right)$ with the radiation of a
phonon of energy $\Delta$  \cite{universal}. Its generalization to
finite temperature is $\Gamma_e = \Gamma_0\coth[\Delta/(2k_B T)]$.
At $k_B T \gg \Delta$ it gives $\Gamma_e \propto T$ as in Eq.\
(\ref{rate-rigid}) obtained for the rigid body. Comparison of the
decoherence provided by the two models will be done in Section
\ref{sec-conclusion}.

As is clear from the derivation, the above result corresponds to
the decoherence of a weakly excited state of the flux qubit. Our
method, however, permits study of decoherence of the state
prepared with ${\bf u} = 0$ and arbitrary $\sigma_z$ (including
$\sigma_z = 1$) at $t = 0$. Dynamics of the vector ${\bm \sigma}$
consists of fast precession about the $X$-axis and slow relaxation
towards the energy minimum that according to Eq.\
(\ref{expansion}) corresponds to $\sigma_x = 1$, $\sigma_{y,z}=0$.
It is accompanied by radiation of sound due to the torque acting
on the flux qubit from the oscillating current. Noticing that the
space-time Fourier transform of the displacement generated by the
torque, ${\bf u}({\bf k},\omega)$, and the time Fourier transform,
${\bm \sigma}(\omega)$, of ${\bm \sigma}(t)$ are always related by
Eqs.\ (\ref{u-k}) due to the linearity of Eqs.\ (\ref{ux}) and
(\ref{uy}), one can transform the integral in Eqs.\ (\ref{LLx})
and (\ref{LLz}) as
\begin{equation}\label{u-int}
\int d^3 r \delta({\bf r})\left(\frac{\partial u_x}{\partial y} -
\frac{\partial u_y}{\partial x}\right) = \frac{l\Delta}{6\pi
v_t^5}\int \frac{d\omega}{2\pi}i\omega^3\sigma_y(\omega)e^{-
i\omega t}\,.
\end{equation}
To the first approximation, fast-precessing and slowly-relaxing
solution of Eqs.\ (\ref{LLy}) and (\ref{LLz}) that satisfies
$\sigma_x^2 + \sigma_y^2 + \sigma_z^2 = 1$ is
\begin{eqnarray}\label{sigma-xy}
\sigma_y(t) & = &
\sqrt{1-\langle\sigma_x\rangle^2}
\sin\left(\frac{\Delta}{\hbar}t\right) \nonumber \\
\sigma_z(t) & = &
\sqrt{1-\langle\sigma_x\rangle^2}
\cos\left(\frac{\Delta}{\hbar}t\right)\,,
\end{eqnarray}
where $\langle\sigma_x\rangle$ is a slow function of time. Within
this approximation the Fourier transform of $\sigma_y$ in Eq.\
(\ref{u-int}) is dominated by the Fourier transform of
$\sin(t\Delta/\hbar)$ that equals
\begin{equation}
i\pi[\delta(\omega + \Delta/\hbar) - \delta(\omega
-\Delta/\hbar)]\,,
\end{equation}
so that the integral (\ref{u-int}) becomes
\begin{equation}\label{int-final}
-2\frac{\hbar\Gamma_0}{\Delta}\sqrt{1-\langle\sigma_x\rangle^2}
\cos\left(\frac{\Delta}{\hbar}t\right)
\end{equation}
where $\Gamma_0$ is given by Eq.\ (\ref{Gamma}). Substituting this
result into Eq.\ (\ref{LLx}), taking into account the first of
Eqs.\ (\ref{sigma-xy}), and averaging the resulting equation over
fast oscillations, $\langle\cos^2(t\Delta/\hbar)\rangle = 1/2$,
one obtains
\begin{equation}
\frac{\partial \langle\sigma_x\rangle}{\partial t} = \Gamma_0
\left(1-\langle\sigma_x\rangle^2\right)\,.
\end{equation}
This leads to the following relaxation law at $t > 0$ after the
system was prepared in the state with arbitrary $\sigma_x =
\tanh(\Gamma_0 t_0) \leq 1$ at the moment of time $t = 0$:
\begin{eqnarray}
&& \langle \sigma_x \rangle = \tanh[\Gamma_0 (t + t_0)] \label{sigma-x}\\
&& \sigma_y =
\frac{{\sin}\left(\frac{\Delta}{\hbar}t\right)}{\cosh[\Gamma_0
(t+t_0)]}
\label{sigma-y}\\
&& \sigma_z =
\frac{{\cos}\left(\frac{\Delta}{\hbar}t\right)}{\cosh[\Gamma_0
(t+t_0)]}\,. \label{sigma-z}
\end{eqnarray}

Our previous consideration of small oscillations of $\sigma_{y,z}$
(that is, precession around $\sigma_x \rightarrow 1$) corresponds
to the choice of $\Gamma_0 t_0 \gg 1$, in which case the decay of
the oscillations is always exponential with the rate $\Gamma_0$,
as has been previously found. If the system is prepared in the
state with $\sigma_z = 1$ (that corresponds to the choice of $t_0
= 0$ in the above equations), it exhibits exponential relaxation,
\begin{equation}
\sigma_z = 2 e^{-\Gamma_0 t}\cos(t\Delta/\hbar)\,,
\end{equation}
only at $\Gamma_0 t \gg 1$. The initial relaxation at $\Gamma_0 t
\ll 1$ is slower:
\begin{equation}
\sigma_z = \frac{\cos(t\Delta/\hbar)}{1+ \frac{1}{2}(\Gamma_0
t)^2}\,.
\end{equation}
This later result for a two-state system should be taken with a
grain of salt, though, as it is likely to be the consequence of
the approximation in which the expectation value of the second
term in Eq.\ (\ref{expansion}) is replaced by the product of
expectation values of ${\sigma}_y$ and phonon field. Such
approximation neglects quantum correlations between spin 1/2 and
the boson field. In this connection, it is interesting to notice
that our model can be easily extended to a system of more than one
flux qubit if all the qubits have the same resonance frequency,
$\omega = \Delta/\hbar$, and are located within a distance from
each other that is small compared to the wavelength of sound of
frequency $\omega$. Indeed, for such a system ${\bm \sigma}/2$ in
Eq.\ (\ref{SQUID-phonons}) gets replaced with the total effective
spin ${\bf S} = {\bm \sigma}_1/2 + {\bm \sigma}_2/2 + {\bm
\sigma}_3/2 + ...$. Since the resulting Hamiltonian is linear on
${\bf S}$, it commutes with ${\bf S}^2$. Consequently, when the
number of qubits, $N$, is large, ${\bf S}$ must behave as a
classical large spin of constant length. In this case, the
approximation that neglects quantum correlations must be good. It
leads to the same equations (\ref{ux}) - (\ref{LLz}) in which
${\bm \sigma}$ is replaced with $N{\bm \sigma}$. This amplifies
the amplitude of sound by a factor $N$. Consequently, $\Gamma_0$
is amplified by a factor $N^2$. One immediately recognizes Dicke
superradiance \cite{Dicke} in this effect. We, therefore, expect
Eqs.\ (\ref{sigma-x}) - (\ref{sigma-z}) with $\Gamma_0 \rightarrow
N^2\Gamma_0$ to correctly describe decoherence in a system of $N
\gg 1$ closely packed flux qubits.

\subsection{Renormalization of the tunnel splitting by the elastic
environment \label{sub-renormalization}}

The above consideration shows that decoherence of the flux qubit
in the elastic environment is dominated by phonons of energy
$\Delta$. Meantime, even at $T = 0$ there are zero-point
oscillations of the solid that produce elastic twists. Such twists
interact with the flux qubit and, as we shall see below,
renormalize the tunnel splitting. This problem cannot be treated
semiclassically as it requires consideration of the entanglement
of the qubit with the excitation modes of the solid. It is based
upon computation of the quantum average of the Hamiltonian
(\ref{SQUID-phonons}), $\langle 0|\hat{H}|0\rangle$ over the
ground state of the solid, $|0\rangle$, that has no real phonons.

Noticing that
\begin{eqnarray}
&& \langle 0|e^{l\xi_{{\mathbf{k}\lambda }}\left(
a_{\mathbf{k}\lambda }-a_{\mathbf{k}\lambda }^{\dagger }\right)}|0
\rangle = \langle 0|e^{-l\xi_{{\mathbf{k}\lambda }}\left(
a_{\mathbf{k}\lambda }-a_{\mathbf{k}\lambda }^{\dagger }\right)}|0
\rangle  \nonumber \\
&& = 1 - \frac{1}{2}|l\xi_{{\mathbf{k}\lambda }}|^2 + ... =
e^{-|l\xi_{{\mathbf{k}\lambda }}|^2/2}\,,
\end{eqnarray}
one obtains
\begin{equation}
\hat{H}_{\sigma} \equiv  \langle 0|\hat{H}|0\rangle =
-\frac{\Delta}{2}\exp\left(-\frac{l^2}{2}\sum_{\mathbf{k}\lambda
}|\xi_{{\mathbf{k}\lambda }}|^2\right)(\sigma_+ + \sigma_-) \,,
\end{equation}
that is,
\begin{equation}
\hat{H}_{\sigma} = -\frac{\Delta_{\rm eff}}{2}\sigma_x\,,
\end{equation}
where
\begin{equation}\label{eff}
\Delta_{\rm eff} = \Delta
\exp\left(-\frac{l^2}{2}\sum_{\mathbf{k}\lambda
}|\xi_{{\mathbf{k}\lambda }}|^2\right)
\end{equation}
is the tunnel splitting renormalized by zero-point quantum elastic
twists. Here $\xi_{{\mathbf{k}\lambda }}$ is given by Eq.\
(\ref{xi}).

The sum over ${\bf k}$ in Eq.\ (\ref{eff}) can be computed by
replacing it with the integral $V\int d^3k/(2\pi)^3$. For the two
transverse phonon modes ${\bf k}\times{\bf e}_{{\bf k}t_1} = \pm k
{\bf e}_{{\bf k}t_2}$. Averaging over the angles then gives
$\langle[{\bf e}_{{\bf k}t}]_z^2\rangle=1/3$. Integrating over $k$
from zero to $k_{\rm max}$ determined by the size of the flux
qubit, one obtains
\begin{equation}\label{eff-kmax}
\Delta_{\rm eff} = {\Delta}\exp\left(-\frac{\hbar l^2k_{\rm
max}^4}{48\pi^2\rho v_t}\right)
\end{equation}
A quick estimate (see Section \ref{sec-conclusion}) shows that the
exponent in Eq.\ (\ref{eff-kmax}) is always small, thus providing
negligible renormalization of the tunnel splitting in a flux
qubit. However, the above result illustrates an important point.
If, for some reason, the shear modulus of the solid, $G = \rho
v_t^2$, disappeared, this, according to Eq.\ (\ref{eff-kmax}),
would lead to the disappearence of the tunnel splitting as well.
The latter is a consequence of the conservation of angular
momentum: The current cannot reverse direction if it cannot
transfer momentum to the body. As is discussed in the next Section
this effect may, in principle, be observed in some two-state
systems.

\section{Discussion and Conclusions \label{sec-conclusion}}

We have studied two models that take into account mechanical
effects associated with quantum oscillations of a superconducting
current in a flux qubit. These effects have simple physical
origin. To change direction, the current must transfer momentum to
the underlying crystal lattice. For the current oscillating in a
SQUID loop, it is a microscopic analogue of the Einstein - de Haas
effect: The change in the angular momentum of the current
associated with its magnetic moment must be compensated by the
change in the angular momentum of the body containing the current.
This inevitably entangles quantum states of a flux qubit with
quantum states of a macroscopic body containing the qubit. One can
naively imagine that almost instantaneous decoherence of quantum
states of the macroscopic body would have a detrimental effect on
the decoherence of the flux qubit. We show that this is not the
case due to the selection rule originating from conservation of
angular momentum. While quantum state of a macroscopic system is,
in general, an admixture of a large number of rotational states
corresponding to different total angular momenta, only tunnel
splittings of the states belonging to the same $J$ contribute to
quantum oscillations of the superconducting current. Broadening of
the tunnel splitting by the rotational states of a qubit is small
as long as the body is sufficiently large.

In the first part of the paper we have studied an exactly solvable
model of a flux qubit entangled with a rigid mechanical rotator.
We show that decoherence in such a system is weak due to inverse
proportionality of the decoherence rate, $\Gamma_r = ({4\hbar
l^2}/{I})({k_B T}/{\Delta})$, to the moment of inertia of the
rotator, $I$. To put things in perspective, consider, e.g., a
micron-size flux qubit embedded in a body of a comparable small
size that is free to rotate. Sound of frequency $\omega =
\Delta/\hbar \sim 10^{10}$s$^{-1}$ would have a wavelength
comparable to the size of the body. Consequently, in reaction to
the oscillations of the superconducting current, such a system
would rotate as a whole, making the rigid-body approximation
developed in Section \ref{sec-rigid} a reasonably good one.
Typical value of the moment of inertia of a micron-size body is in
the ballpark of $10^{-19}$g$\cdot$cm$^2$. Taking $l \sim 10^5$ for
a micron-size current loop, one obtains the following values of
the parameters in equations (\ref{beta-T}) - (\ref{rate-rigid}):
$\beta_l \approx 2\times10^{-8}$, $\beta_T \approx 3\times
10^{-4}(k_B T/\Delta)^{1/2}$, $\sigma_{\infty} \sim
4\times10^{-8}(k_B T/\Delta)$. Decoherence is dominated by $J \sim
10^9(k_B T/\Delta)^{1/2}$, which corresponds to frequencies of the
rotational Brownian motion $\omega = \hbar J/I \sim 10(k_B
T/\Delta)^{1/2}$s$^{-1}$. This provides $\Gamma \sim 500$s$^{-1}$
that corresponds to a rather high quality factor of quantum
oscillations, $Q = \Delta/(\hbar\Gamma) \sim 2\times
10^7[\Delta/(k_BT)]$, even in the extreme case of a micron size
system.

In the second part of the paper we have studied interaction of the
flux qubit with the twists of the elastic body, dictated by the
conservation of angular momentum. Such model has no free
parameters. While its exact quantum states are not known, one can
develop a reasonably good approximation in which the internal
torque produced inside the body by the oscillating current is
treated as a source of elastic shear waves. If the elastic
environment is considered to be infinite in space, this is an open
system as compared to the closed system that consists of a
finite-size rotator with a flux qubit. In the infinite elastic
system the shear waves generated by the point source of torque
escape to infinity, thus allowing finite decoherence at $T = 0$ as
compared to the closed system. The corresponding decoherence rate
is given by $\Gamma_e = {l^2\Delta^5}/({12\pi \hbar^4 \rho
v_t^5})\coth[\Delta/(2k_B T)]$. At $l \sim 10^5$, $\omega =
\Delta/\hbar \sim 10^{10}$s$^{-1}$ it is of the order of
$10^6$s$^{-1}$, which provides $Q = \Delta/(\hbar\Gamma) \sim
10^4$. This shows that the effect studied in this paper, while
allowing weak decoherence, can hardly be ignored in designing flux
qubits.

A good check of the validity of the above results can be obtained
by comparing decoherence rates obtained within the rigid-rotator
model and within the elastic model. At $k_B T \geq \Delta$ the
ratio of the two rates is $\Gamma_{e}/\Gamma_{r} = (4\pi^4/3)
(I/\rho \lambda^5)$ where $\lambda = 2\pi \hbar v_t/\Delta$ is the
wavelength of shear waves of frequency $\omega = \Delta/\hbar$.
Noticing that the moment of inertia of a rigid body of radius $R$
is of order $\rho R^5$, we see that $\Gamma_{e}/\Gamma_{r} \sim 1$
at $\lambda \sim 2R$. This agreement between the two models that
consider the same effect from two very different angles is quite
remarkable.

In our consideration of the conservation of angular momentum,
certain effects that may exist in real systems have been left out.
Among them are interactions of the flux qubit with magnetic atoms
and nuclear spins that can, in principle, absorb some part of the
angular momentum of the SQUID. For $l \gg 1$ such processes must
be suppressed, however, as they require coherent participation of
many magnetic atoms and many nuclear spins. Interaction of the
flux qubit with the shear waves of the body must be the primary
mechanism of the conservation of angular momentum. Being
unavoidable, it imposes a universal upper bound on the quality
factor of the qubit.

The effect of rotations on decoherence can also be understood from
another angle. At $\phi = \omega t$ that corresponds to the
uniform rotation of the flux qubit about the $Z$-axis the
Hamiltonian (\ref{projection'}) is equivalent to the Hamiltonian
of spin 1/2 in the effective magnetic field of amplitude
$\Delta/(2\mu_B)$ ($\mu_B$ being the Bohr magneton) rotating in
the $XY$ plane at an angular velocity $\Omega = 2l\omega$.
Switching to the coordinate frame rotating with the field, gives
an effective constant field applied along the X-axis plus the
effective bias field in the Z-direction, $\hat{H}_{\sigma}'' =
-l\hbar\omega\sigma_z - \frac{\Delta}{2}\sigma_x$. The first term
is simply $-{\bm \omega}\cdot \hbar{\bf l}$, that appears in the
frame rotating at the mechanical angular velocity ${\bm \omega}$,
projected into the $|\uparrow\rangle$ and $|\downarrow\rangle$
states. Real bias magnetic field $B$ adds the term $-{\bf
B}\cdot(\mu_B {\bf l})$ to the Hamiltonian. When the field is
applied along the Z-axis the full two-state Hamiltonian in the
rotating (SQUID) frame of reference becomes $\hat{H}_{\sigma}''
=-l(\hbar\omega + \mu_B B)\sigma_z - \frac{\Delta}{2}\sigma_x$.
This proves that the rotation of a truncated two-state SQUID
system satisfies Larmor theorem. It is equivalent to the magnetic
field $B/\omega = \hbar/\mu_B \sim 10^{-7}$Oe/Hz. Effective fields
generated by slow rotations of the equipment must have negligible
effect on the flux qubit. However, the effect of local dynamic
shear deformations on a microscopic SQUID must be noticeable
because the corresponding angular velocities $(ku)(\Delta/\hbar)$
can easily reach $10^7$Hz, providing effective fields in the range
of $1G$.

Experiments with flux qubits have shown that significant
decoherence comes from $1/f$ noise, the origin of which has been
debated \cite{Yoshihara-PRL06,Koch-PRL07}. Notice in this
connection that relaxation of microscopic shear strains in a solid
must be a source of dynamical local twists that, according to the
above discussion, generate local effective magnetic fields. It is,
therefore, plausible that relaxation of shear strains at the
location of the qubit is, in fact, responsible for the observed
$1/f$ noise affecting quantum dynamics of the qubit.

Another observation worth mentioning is amplification of
decoherence in a system of flux qubits positioned in close
proximity to each other. This effect may be important in designing
architectures of flux qubits if they are to be used for quantum
computing. It will reveal itself when $N$ microscopic qubits with
identical tunnel splitting $\Delta$ are positioned within the
wavelength of sound of frequency $\Delta/\hbar$. As has been
demonstrated in Section \ref{sub-decoherence-elastic}, radiation
of sound by such a system and, thus, decoherence will be amplified
by a factor $N^2$. This is an acoustic analogue of Dicke
superradiance that may impose an upper limit on the density of
flux qubits. One way to avoid this effect in a dense assembly of
qubits would be to use qubits of significantly different $\Delta$.

In Section \ref{sub-renormalization} we studied renormalization of
the tunnel splitting of a flux qubit arising from its interaction
with zero-point shear deformations. The magnetic moment of the
current of strength $J$ in a loop of area $a$ is $\mu = Ja/c$,
which gives $l = Ja/(c\mu_B)$. With $a=\pi r^2$ and $k_{\rm max} =
2\pi/r$, the exponent in Eq.\ (\ref{eff-kmax}) becomes $\pi^4\hbar
J/(3c^2\mu_B^2\rho v_t)$. At $J \sim 1\mu$A it is hopelessly
small, thus, making this kind of renormalization irrelevant for a
flux qubit. Notice in this connection that a similar effect,
described by Eq.\ (\ref{eff-kmax}), may exist for the tunnel
splitting of the atomic magnetic cluster. In this case $l$ would
be significantly smaller but $k_{\rm max}$ would be much greater
than for a flux qubit. An estimate for, e.g., a magnetic molecule
frozen in solid He-4 shows that the exponent in Eq.\
(\ref{eff-kmax}) can easily be of order unity. As the He-solid
approaches melting transition on decreasing pressure, its shear
modulus would go to zero, resulting in the freezing of tunneling.

Finally, we would like to notice that treatment developed in this
paper should apply to nanomechanical devices incorporating SQUIDs.
Such devices have been recently made and measured
\cite{LaHaye-Nat09,Connell-Nat10}. They open the whole new field
of the entanglement of qubit states with mechanical oscillations.
Possible manipulation of superconductng qubits by mechanical
rotations is another interesting aspect of the research on
nanomechanical superconducting qubits. Our model of a rigid
rotator with a flux qubit may provide a framework for theoretical
studies of these effects.

\section{Acknowledgements}

This work has been supported by the U.S. Department of Energy
Grant No. DE-FG02-93ER45487.

\end{document}